\documentstyle[12pt,amssymb,bezier]{article}
\begin{document}
\newcommand{\be}{\begin{equation}}
\newcommand{\ee}{\end{equation}}
\newcommand{\ba}{\begin{eqnarray}}
\newcommand{\ea}{\end{eqnarray}}
\newcommand{\by}{\begin{eqnarray*}}
\newcommand{\ey}{\end{eqnarray*}}
\newcommand{\vp}{\varphi}
\newcommand{\e}{\epsilon}
\newcommand{\ve}{\varepsilon}
\newcommand{\p}{\partial}
\newcommand{\ra}{\rightarrow}
\newcommand{\La}{\Lambda}
\newcommand{\la}{\lambda}
\newcommand{\Om}{\Omega}
\newcommand{\om}{\omega}
\newcommand{\al}{\alpha}
\newcommand{\Del}{\Delta}
\newcommand{\del}{\delta}
\newcommand{\si}{\sigma}
\newcommand{\Si}{\Sigma}
\newcommand{\f}{\frac}
\newcommand{\ti}{\tilde}
\newcommand{\pr}{\prime}
\newcommand{\ct}{\cite}
\newcommand{\ga}{\gamma}
\newcommand{\Ga}{\Gamma}
\newcommand{\scs}{\scriptstyle}
\newcommand{\sr}{\stackrel}
\newcommand{\ts}{\times}
\newcommand{\ul}{\underline}
\newcommand{\nn}{\nonumber}
\hfill Preprint IHEP 98-56
\begin{center}
{\bf Magnetic Catalysis and Oscillating Effects in the Nambu --
Jona-Lasinio Model at Nonzero Chemical Potential \footnote{Poster talk 
given at the "5th International 
Workshop on Thermal Field Theories and Their Application", 10-14 August 1998,
Regensburg, Germany.}}

\vspace*{0.4cm}
K.G. Klimenko \footnote{E-mail:~kklim@mx.ihep.su}

\vspace{0.2cm} 
Institute for High Energy Physics, 142284 Protvino, Moscow region, Russia

\end{center}

\vspace{0.4cm}

Phase structure of the four dimensional Nambu -- Jona-Lasinio
model has been investigated in two cases: 1) in nonsimply
connected space-time of the form $R^3\times S^1$ (space
coordinate is compactified and the length of the circle $S^1$ is
$L$) with nonzero chemical potential $\mu$ and 2) in Minkowski
space-time at nonzero values of $\mu,H$, where $H$ is the
external magnetic field. In both cases on phase portraits of the
model there are infinitly many massless chirally symmetric
phases as well as massive ones with spontaneously broken chiral
invariance. Such phase structure leads unavoidably to oscillations
of some physical parameters at $L\to\infty$ or $H\to 0$,
including magnetization, pressure and particle density of the system
as well as quark condensate and critical curve of chiral phase transitions.
Phase transitions of 1st and 2nd orders and several tricritical
points have been shown to exist on phase diagrams of the model.

\vspace*{0.6cm}

\section*{1. Introduction
\footnote{This report is based on works done in
collaboration with A.K. Klimenko, M.A. Vdovichenko, A.S. Vshivtsev and
V.Ch. Zhukovskii \ct{1,2}.}}

The concept of dynamical chiral symmetry breaking (DCSB) plays
an essential role in elementary particle physics and quantum
field theory (QFT). In QFT this phenomenon is well observed in
Nambu -- Jona-Lasinio (NJL) type models -- four-dimensional models
with four-fermionic interactions \ct{nam,vaks}. The simplest one 
is presented by the Lagrangian
\be
L_\psi=\sum^{N}_{k=1}\bar\psi_k i\hat\partial\psi_k
+\frac{G}{2N}
[(\sum^{N}_{k=1}\bar\psi_k\psi_k)^2
+(\sum^{N}_{k=1}\bar\psi_k i\gamma_5\psi_k)^2],
\ee
which is invariant under continuous chiral transformations
\be
\psi_k\to e^{i\theta\gamma_5}\psi_k~~;~~ (k=1,...,N).
\ee
(In order to apply large N-expansion technique we use here N-fermionic version
of the model.)

Since there are no closed physical systems in nature, the influence of
different external factors on the DCSB mechanism is of great interest.
In these realms, special attentions have been given to analysis of the
vacuum structure of the NJL type models at nonzero temperature and chemical
potential \ct{kaw,1}, in the presence of external (chromo-)magnetic fields
\ct{klev,eb,gus}, with allowance for curvature and nontrivial space-time
topology \ct{In,2}. Combined action of external electromagnetic
and gravitational fields on DCSB effect in four-fermion field theories
were investigated in \ct{muta,od}.

In the present paper we consider the phase structure and related
oscillating effects of the four dimensional Nambu --
Jona-Lasinio model in two cases: 1) in nonsimply connected
space-time of the form $R^3\times S^1$ (space coordinate is
compactified) with nonzero chemical potential $\mu$ and 2) in
Minkowski space-time at nonzero values of $\mu,H$, where $H$ is
the external magnetic field.

\subsubsection*{1.1 NJL model at $\mu\neq 0$.}

First of all let us prepare the basis for investigations in following
sections and consider phase structure of the model (1) at $\mu\neq 0$ in
Minkowski space-time.

Recall some well - known vacuum properties
of the theory (1) at $\mu=0$. The introduction of an auxiliary
Lagrangian
\be
\tilde L=\bar\psi i\hat\partial\psi-
\bar\psi (\sigma_1+i\sigma_2\gamma_5)
\psi-\frac{N}{2G}(\sigma^2_1+\sigma^2_2)
\ee
greatly facilitates the problem under consideration. (In (3) and
other formulae below we have omitted the fermionic index $k$ for
simplicity.) On the equations of motion for auxiliary bosonic
fields $\sigma_{1,2}$ the theory (3) is equivalent to the (1) one.

From (3) it follows in the leading order of $1/N$ - expansion:
$$
\exp (iN S_{eff}(\sigma_{1,2})) =
\int D\bar\psi D\psi \exp(i\int\tilde L d^4x),
$$
where
\[
S_{eff}(\sigma_{1,2})=-
\int d^4x\frac{\sigma_1+\sigma^2_2}{2G}-i\ln\det
(i\hat\partial-\sigma_1-i\gamma_5\sigma_2).
\]
Supposing that in this formula $\sigma_{1,2}$ are not dependent from 
the space - time points we have by definition:
\[
S_{eff}(\sigma_{1,2})=-
V_{0}(\sigma_{1,2})\int d^4 x,
\]
where $(\Sigma=\sqrt{\sigma^2_1+\sigma^2_2})$:
\be
V_{0}(\sigma_{1,2})=
\frac{\Sigma^2}{2G}+2i
\int \frac{d^4p}{(2\pi)^4}\ln
(\Sigma^2-p^2)\equiv V_0(\Sigma).
\ee

Introducing in (4) Euclidean metrics $(p_0\to ip_0)$ and cutting off the range
of integration $(p^2\leq \Lambda^2)$, we obtain:
\be
V_0(\Sigma)=
\frac{\Sigma^2}{2G}-
\frac{1}{16\pi^2}\Biggl\{
\Lambda^4\ln
\left(1+\frac{\Sigma^2}{\Lambda^2}\right)+\Lambda^2\Sigma^2-\Biggr.\nonumber \\
-\Sigma^4\ln \left(1+\frac{\Lambda^2}{\Sigma^2}\right)\Biggl.\Biggr\}.
\ee
The stationary equation for the effective potential (5) has the form:
\be
\frac{\partial V_0(\Sigma)}{\partial\Sigma}=0=
\frac{\Sigma}{4\pi^2}
\Biggl\{\frac{4\pi^2}{G}-\Lambda^2+\Sigma^2 \ln
\left (1+\frac{\Lambda^2}{\Sigma^2}\right)\Biggr\}\equiv
\frac{\Sigma}{4\pi^2}F(\Sigma).
\ee
Now one can easily see that at $G<G_c=4\pi^2/\Lambda^2$ eq. (6) has no 
solutions apart from $\Sigma=0$. Hence, in this case fermions are massless,
and chiral invariance (2) is not broken.

If $G>G_c$, then Eq. (6) has one nontrivial solution
$\Sigma_0(G,\Lambda) \not = 0$ such that $F(\Sigma_0)=0$. In
this case $\Sigma_0$ is a point of global minimum for the
potential $V_0(\Sigma)$. This mean that spontaneous breaking of
the symmetry (2) takes place. Moreover, fermions acquire mass
$M\equiv\Sigma_0 (G,\Lambda)$.

Let us now imagine that $\mu>0$ and temperature $T \not =0$. In
this case one can find effective potential $V_{\mu T}(\Sigma)$
if the mesure of integration in (4) is transformed according to a rule
$$
\int\frac{dp_0}{2\pi}\to iT\sum^{\infty}_{n=-\infty},\,\,
p_0\to i\pi T(2n+1)+\mu.
$$
Summing there over $n$ \ct{dol} and directing in the obtained
expression temperature to zero, we have:
\be
V_{\mu}(\Sigma)=V_0(\Sigma) -
2\int\f{d^3p}{(2\pi)^3}\theta(\mu - \sqrt{\Sigma^2+p^2})(\mu -
\sqrt{\Sigma^2+p^2}), 
\ee
where $\theta(x)$ is the step function. Integrating in (7), we find
\ba
V_{\mu}(\Sigma)=V_0(\Sigma)&-&
\frac{\theta(\mu-\Sigma)}{16\pi^2}
\Biggl\{\frac{10}{3}\mu(\mu^2-\Sigma^2)^{3/2}- \Biggr.\nonumber \\
-2\mu^3\sqrt{\mu^2-\Sigma^2}&+&\Sigma^4\ln
~[~(~\mu+\sqrt{\mu^2-\Sigma^2}~)^2/\Sigma^2~]\Biggl.\Biggr\}.
\ea
\begin{center}
\unitlength=1mm
\special{em:linewidth 0.4pt}
\linethickness{0.4pt}
\begin{picture}(145.67,88.33)
\put(11.00,8.33){\vector(0,1){80.00}}
\put(4.67,17.33){\vector(1,0){141.00}}
\linethickness{1.0pt}
\put(11.00,17.33){\line(2,1){80.33}}
\put(59.67,63.00){\line(5,1){7.67}}
\put(71.67,64.67){\line(5,-1){8.67}}
\put(83.33,61.33){\line(2,-1){10.00}}
\put(95.67,55.00){\line(4,-1){9.67}}
\put(108.67,52.00){\line(6,1){11.00}}
\put(122.00,54.67){\line(3,2){8.00}}
\bezier{336}(11.00,17.33)(26.67,65.00)(60.00,63.00)
\put(60.00,17.33){\line(0,1){2.33}}
\put(91.00,17.33){\line(0,1){2.33}}
\put(140.33,21.33){\makebox(0,0)[cc]{$\scs{M}$}}
\put(16.67,85.33){\makebox(0,0)[cc]{$\mu$}}
\put(84.33,37.67){\makebox(0,0)[cc]{$B$}}
\put(44.33,47.67){\makebox(0,0)[cc]{$C$}}
\put(60.00,63.00){\circle*{1.33}}
\put(60.00,68.00){\makebox(0,0)[cc]{$\alpha$}}
\put(92.00,61.67){\makebox(0,0)[cc]{$\beta$}}
\put(91.00,57.00){\circle*{1.33}}
\put(37.67,80.00){\makebox(0,0)[cb]{$A$}}
\put(21.67,50.67){\makebox(0,0)[cc]{$\mu_{1c}$}}
\put(53.00,33.00){\makebox(0,0)[cc]{$\mu_{2c}$}}
\put(124.67,62.00){\makebox(0,0)[cc]{$\mu_{3c}$}}
\put(60.00,13.00){\makebox(0,0)[cc]{$M_{1c}$}}
\put(91.00,13.00){\makebox(0,0)[cc]{$M_{2c}$}}
\end{picture}
\end{center}
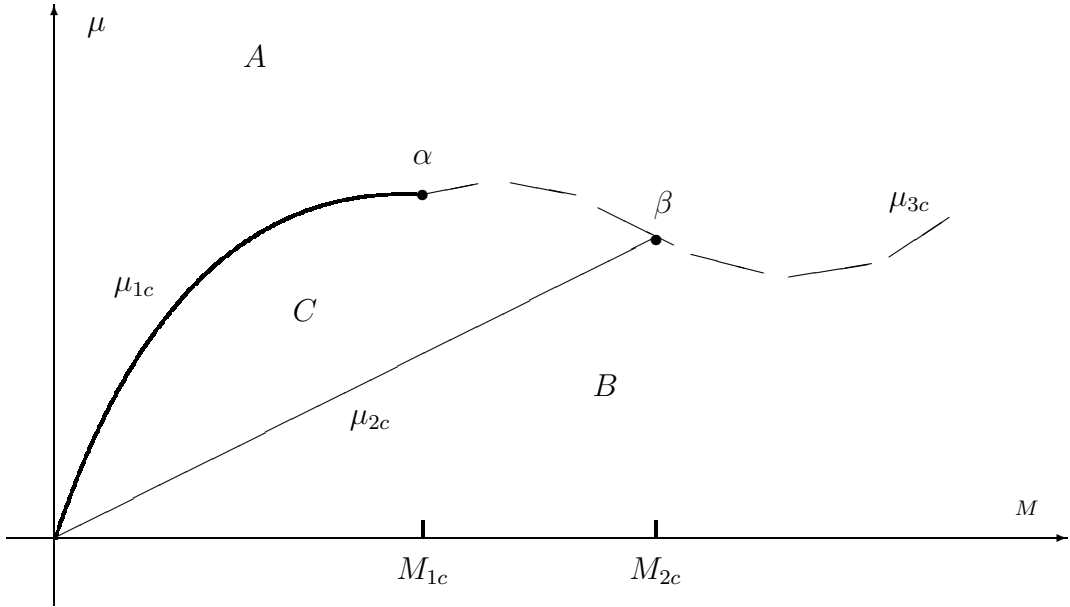
{\scriptsize FIG.1 Phase portrait of the NJL model at nonzero $\mu$ and
for arbitrary values of fermionic mass $M$. Phases $B$ and $C$ are 
massive nonsimmetric phases, $A$ is chirally simmetric phase. Here $\mu_{2c}=M$,
$\mu_{1c}=\sqrt{\f 12 M^2\ln(1+\La^2/M^2)}$, $M_{2c}=\La/(2.21...)$,
$M_{1c}$ is the solution of the equation $\mu_{1c}^2(M_{1c})=\La^2/(4e)$.
In the phase $B$ particle density in the ground state is equal to zero.
However, in phase $C$ particle density is not zero.}
\vspace{0.5cm}

It follows from (8) that in the case $G<G_c$ and at arbitrary
values of chemical potential chiral symmetry (2) is not broken.
However, at $G>G_c$ the model has a rich phase structure, which
is presented in the Fig.1 in terms of $\mu$ and $M$. (At $G>G_c$
one can use the fermionic mass $M$ as an independent parameter
of the theory. Three quantities $G$, $M$ and $\Lambda$ are
connected by the Eq. (6).) In this Figure the solid and dashed
lines represent the critical curves of the second- and
first-order phase transitions, respectively.  Futhermore, there
are two tricritical points $\al$ and $\beta$, two massive phases
$B$ and $C$ with spontaneously broken chiral invariance as well
as the simmetric massless phase $A$ on the phase portrait of the
NJL model (detailed calculations of the vacuum
structure of the NJL model one can find in \ct{1}).

\section*{2. Phase structure of the NJL model at $\mu\neq 0$ and in the
$R^3\times S^1$ space-time}

It is well-known that unified theory of all forces (including
gravitation) of the nature is not constructed up to now.  Since
in early Universe the gravity was sufficiently strong, so one
should take it into account, many physicists study quantum field
theories in space-times with
nontrivial metric and topology.  At this, NJL model is the
object of special attention (see the review \ct{muta}), because the idea of
dynamical chiral symmetry breaking is the underlying concept in
elementary particle physics. There is a copious literature on this subject
\ct{In,muta,od,kim}. In particular, the investigation of four-fermion
theories in the space-time of the form $R^d\ts S^1\ts\cdots\ts S^1$ is 
of great interest \ct{kim}. The matter is that such space-time topology
occurs in superstring theories, in description of Casimir type effects
and so on.

In the present section the NJL model in the $R^3\times S^1$ 
space-time and at $\mu\neq 0$ is considered since great amount of 
physical phemomena take place at nonzero particle density, i.e. at nonzero
chemical potential. Here space coordinate is
compactified and the circumference $S^1$ has the length $L$. For simplisity we 
study only the case with periodic boundary conditions:
$\psi(t,x+L,y,z) = \psi(t,x,y,z).$

\subsubsection*{2.1 Phase structure}

In order to find the effective potential $V_{\mu L}(\Si)$ at $\mu\neq 0$ and
$L\not=\infty$, we need to transform the integration over $p_1$ in (7) into
a summation over discrete values $p_{1n}$ according to a rule
\[
\int\frac{dp_1}{2\pi}f(p_1)\rightarrow \frac{1}{L}\sum^{\infty}_{n=-\infty}
f(p_{1n});~~~p_{1n} = 2\pi n/L~, ~n=0,\pm 1,\pm 2,...
\]
The resulting expression is
\[
V_{\mu L} (\Sigma)=V_L (\Sigma)-\f{\la}{6\pi}
\sum^{\infty}_{n=0}\al_n
\theta (\mu - \sqrt{\Sigma^2 + (2\pi\lambda n)^2}~)\cdot
\]
\be
\cdot (\mu - \sqrt{\Sigma^2 + (2\pi\lambda n)^2}~)^2(\mu
+2\sqrt{\Sigma^2 + (2\pi\lambda n)^2}~),
\ee
where
\be
V_L (\Sigma)=V_0(\Sigma)
-\frac{2}{\pi^2 L}\int\limits^{\infty}_{0}
dxx^2\ln[\;1 - \exp (-L\sqrt{x^2+\Sigma^2}\;)\;],
\ee
$\al_n=2-\delta_{n0}$ and $V_0(\Si)$ is given in (5). The stationary equation
for the function (9) has the form
\ba
\f{\p V_{\mu L} (\Si)}{\p\Si}&=&\f{\p V_L (\Si)}{\p\Si}+
\f{\la\Si}\pi\sum^{\infty}_{n=0}\al_n
\theta (\mu - \sqrt{\Si^2 + (2\pi\lambda n)^2}~)\cdot\nn \\
& &\cdot(\mu - \sqrt{\Si^2 + (2\pi\lambda n)^2}~)
\equiv\f{2\Si}{\pi^2}\phi(\Si)=0.
\ea

\ul{The case $\mu=0,\la\equiv 1/L>0$.} Putting in (9) $\mu$
equals to zero, we obtain effective potential $V_L (\Si)$ (11)
in the case under consideration.  This function at $G>G_c$ has a
global minimum point $\Si_0(\la)>0$, which means that for all
values of $\la\geq 0$ chiral invariance of the model is
spontaneously broken.  Obviously, $\Si_0(\la)\to M$ at
$\la\to 0$.

When $G<G_c$, $\Si_0(\la)\equiv 0$ at $\la<\la_0$ and $\Si_0(\la)> 0$
at $\la>\la_0$ (here $\f{\pi^2}{2G}-\f{\La^2}8\equiv\f{\pi^2}6 \la^2_0$).
In the point $\la=\la_0$ there is a second order phase transition from
a symmetric to a nonsymmetric phase of the model, because
at $\la\to\la_0{\scs +}$
\[
\Si_0(\la) = \f 23\pi(\la-\la_0) + o(\la-\la_0),
\]
i.e. the order parameter $\Si_0(\la)$  is a continuous 
function in the point $\la=\la_0$.
At $\la\to\infty$ for all values of coupling constant $G$ we have
\[
\Si_0(\la) \sim 2\pi\la (2.719...).
\]
Details of above calculations and of following ones are
presented in \ct{2}.

\ul{The general case $\mu,\la\neq 0$.} We shall find a one-to-one
correspondence between points of the plane $(\la,\mu)$ 
and the phase structure of initial model. It is very convinient to divide
this plane into regions $\om_k$:
\be
(\mu,\lambda)=\bigcup\limits^{\infty}_{k=0}\omega_k;\;
\omega_k=\{(\mu,\lambda):
2\pi \lambda k \leq \mu < 2\pi\lambda (k+1)\}.
\ee

In $\om_0$ only the first term from a series in (9) is nonzero, in $\om_1$
only the first and the second terms are nonzero  and so on. In order to
obtain a phase structure one should study
step by step the global minimum point of the function $V_{\mu L}(\Si)$
 in regions $\om_0$,$\om_1$,....
Omiting technical details we show at once resulting
phase portraits at $G_1\equiv (0.917...)G_c<G<G_c$ (see Fig.2) 
and at $G_c<G<(1.225...)G_c\equiv G_2$ (see Fig.3) as well \ct{2}.

One can see in Fig.2 only two massive nonsymmetric phases $B$ and
$C$. In contrast, there are infinitly many massive phases
$C_k (k=0,1,...)$ in Fig.3. In phase $B$ particle density is
identically zero, but in $C$ and in all $C_k$ phases this
quantity is not zero.  In both figures there are also infinitly
many simmetric massless phases $A_k (k=0,1,..)$ of the NJL model.

The line $\mu_0(0)c_nc_2c_1b$ in Fig.3 is the critical line $\mu_c(\la)$ of
the second-order phase transitions, where chiral symmetry is restored.
The $\mu_c(\la)$ is defined by the equation
\be
\phi(0)=0,
\ee
where $\phi(\Si)$ is given in (11). Critical lines $l_1,l_2,...$
are solutions of equations
\[
\phi(\mu_k)\equiv\phi(\sqrt{\mu^2-(2\pi k\la)^2})=0
\]
for $k=1,2,...$ respectively. Boundaries between massless phases in both
figures are boundaries between regions $\om_k$ from (12).

Phase structure of NJL model at another values of coupling constant $G$
is presented in \ct{2}, where one can also find more detailed description of 
above phase portraits at $G_1<G<G_2$.

\begin{center}
\unitlength=1mm
\special{em:linewidth 0.4pt}
\linethickness{0.4pt}
\begin{picture}(144.00,135.00)
\put(16.00,18.00){\vector(1,0){128.33}}
\put(16.00,18.00){\vector(0,1){117.00}}
\linethickness{1.0pt}
\put(16.00,18.00){\line(6,5){97.33}}
\put(16.00,18.00){\line(4,5){71.33}}
\put(16.00,18.00){\line(1,2){49.00}}
\put(16.00,18.00){\line(1,6){17.00}}
\put(89.33,30.67){\line(2,1){4.00}}
\put(96.33,35.00){\line(2,3){3.00}}
\put(101.33,42.67){\line(1,3){1.33}}
\put(104.00,49.67){\line(3,5){2.67}}
\put(108.00,55.67){\line(5,6){3.33}}
\put(113.33,61.33){\line(5,4){5.00}}
\put(120.33,66.33){\line(3,1){5.33}}
\put(128.33,68.67){\line(5,1){7.67}}
\put(87.00,30.00){\circle*{1.33}}
\put(104.00,50.00){\circle*{1.33}}
\put(87.00,27.00){\makebox(0,0)[cc]{$a$}}
\put(107.00,50.00){\makebox(0,0)[cc]{$b$}}
\put(59.33,77.00){\circle*{0.33}}
\put(55.67,78.67){\circle*{0.33}}
\put(51.67,80.33){\circle*{0.33}}
\put(24.67,85.33){\circle*{0.33}}
\put(22.00,85.67){\circle*{0.33}}
\put(18.67,85.67){\circle*{0.33}}
\put(10.00,127.00){\makebox(0,0)[cc]{$\mu$}}
\put(139.33,12.00){\makebox(0,0)[cc]{$\lambda$}}
\put(116.00,29.67){\makebox(0,0)[cc]{$B$}}
\put(73.00,35.67){\makebox(0,0)[cc]{$C$}}
\put(85.67,59.00){\makebox(0,0)[cc]{$A_0$}}
\put(67.33,71.67){\makebox(0,0)[cc]{$A_1$}}
\put(36.33,82.00){\makebox(0,0)[cc]{$A_n$}}
\put(115.00,103.33){\makebox(0,0)[cc]{$\mu=2\pi\lambda$}}
\put(89.33,112.33){\makebox(0,0)[cc]{$\mu=4\pi\lambda$}}
\put(67.00,122.33){\makebox(0,0)[cc]{$\mu=2\pi\lambda n$}}
\put(39.67,13.67){\makebox(0,0)[cc]{$\lambda_0$}}
\linethickness{1.0pt}
\bezier{324}(40.33,18.00)(56.67,48.00)(104.00,50.00)
\bezier{196}(40.33,18.00)(58.67,30.33)(87.00,30.00)
\end{picture}
\end{center}
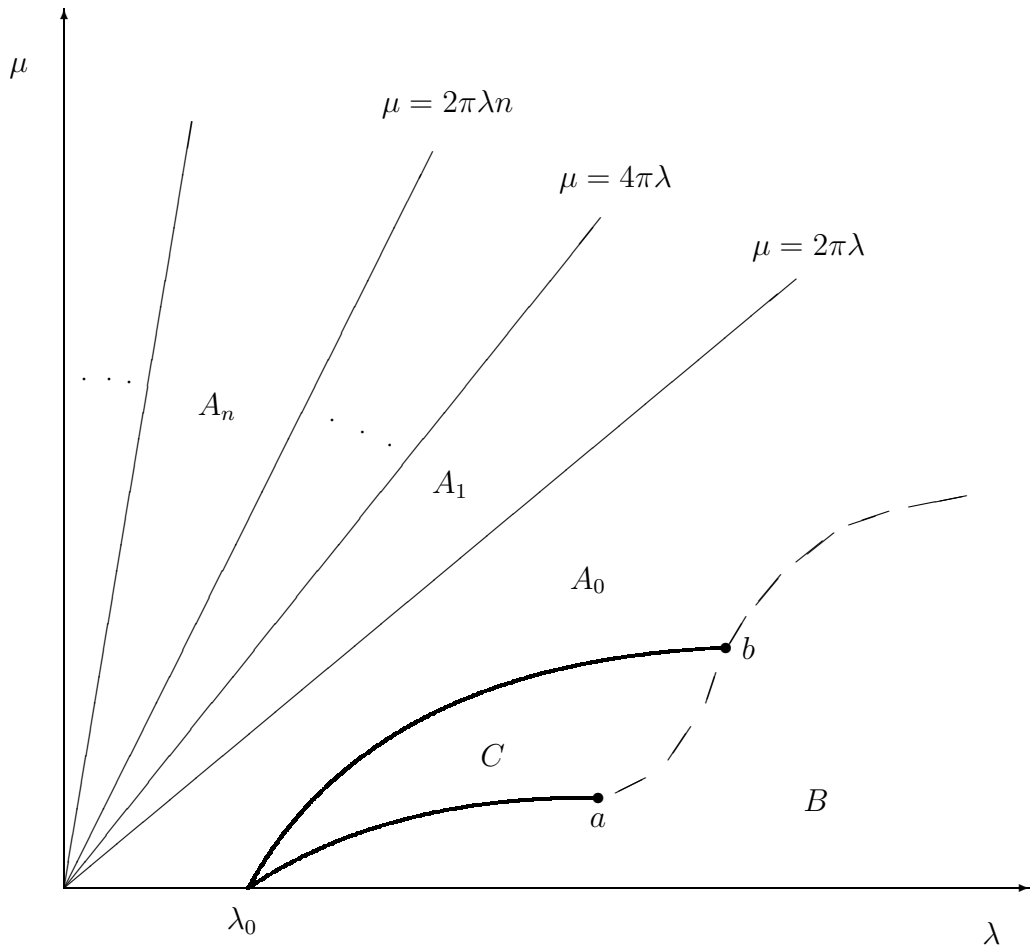
{\scriptsize FIG.2 Phase portrait of the $R^3\ts S^1$ NJL model
at $\mu\neq 0$ and $G_1<G<G_c$ ($\la=1/L$). Dashed lines are critical curves
of first-order phase transition, solid lines correspond to second-order 
critical curves. Points $a$ and $b$ are tricritical ones. There is
a cascade of massless symmetric phases $A_k$ ($k=0,1,2...$).}
\vspace{0.5cm}

\subsubsection*{2.1 Effects of oscillations}

Now let us show that, due to a presence in a phase structure of the NJL
model of cascades of massless $A_k$ as well as massive $C_k$
phases, one can observe oscillations of some physical
parameters. We shall consider only the case $G_c<G<G_2$.

{\bf The continuous physical quantity $f(x)$  is
called as an oscillating one at $x\to a$, if there exist a monotonically
encreasing (decreasing) sequence $\{x_n\}$ such that: i) $x_n\to
a$ at $n\to\infty$, ii) $f(x)$ is a continuous function at each point $x_n$ and
iii) $f'(x)$ is a discontinuous function at points $x_n$.}

At zero temperature oscillating quantity satisfies, as a rule, to this 
definition (see, for example, magnetic oscillations in quantum electrodynamics
\ct{per,as}). Of course, at nonzero temperature one can observe a more
smooth behaviour of oscillating parameters. Since we shall deal with
zero temperature case only, the above cited definition of
oscillations is well suited in the framework of present paper.
\begin{center}
\unitlength=1.00mm
\special{em:linewidth 0.4pt}
\linethickness{0.4pt}
\begin{picture}(148.00,138.00)
\put(14.00,15.00){\vector(0,1){123.00}}
\put(11.67,18.33){\vector(1,0){136.33}}
\linethickness{1.0pt}
\put(90.67,75.67){\line(4,3){43.33}}
\put(63.33,77.00){\line(5,6){33.33}}
\put(44.00,78.00){\line(1,2){21.67}}
\put(29.00,78.33){\line(1,4){11.67}}
\bezier{250}(14.00,32.33)(53.00,32.33)(96.33,42.33)
\bezier{260}(90.67,75.67)(38.00,33.67)(14.00,32.33)
\bezier{200}(63.33,77.00)(29.00,38.00)(14.00,32.33)
\bezier{180}(44.00,78.00)(19.33,36.67)(14.00,32.33)
\bezier{150}(29.00,78.33)(16.67,37.00)(14.00,32.33)
\put(90.67,75.67){\circle*{1.33}}
\put(63.33,77.00){\circle*{1.33}}
\put(44.00,78.00){\circle*{1.33}}
\put(29.00,78.33){\circle*{1.33}}
\bezier{145}(90.67,75.67)(93.33,57.33)(124.33,58.67)
\bezier{150}(90.33,75.67)(61.67,59.33)(63.33,77.00)
\bezier{75}(44.00,78.00)(30.33,67.67)(29.00,78.33)
\put(14.00,74.00){\line(6,-1){4.00}}
\put(19.67,73.33){\circle*{0.33}}
\put(21.67,73.67){\circle*{0.33}}
\put(96.33,42.33){\circle*{1.33}}
\put(124.33,58.67){\circle*{1.33}}
\put(99.00,42.67){\line(6,1){6.00}}
\put(108.00,44.67){\line(2,1){5.67}}
\put(117.00,50.00){\line(1,1){3.67}}
\put(123.00,56.00){\line(3,5){3.33}}
\put(128.33,63.33){\line(1,1){4.00}}
\put(135.00,68.67){\line(5,2){6.33}}
\put(96.67,46.00){\makebox(0,0)[cc]{$a$}}
\put(127.00,55.67){\makebox(0,0)[cc]{$b$}}
\put(119.33,30.67){\makebox(0,0)[cc]{$B$}}
\put(144.00,14.67){\makebox(0,0)[cc]{$\lambda$}}
\put(89.00,78.33){\makebox(0,0)[cc]{$c_1$}}
\put(62.33,79.33){\makebox(0,0)[cc]{$c_2$}}
\put(41.00,79.00){\makebox(0,0)[cc]{$c_n$}}
\put(72.00,48.00){\makebox(0,0)[cc]{$C_0$}}
\put(64.00,61.00){\makebox(0,0)[cc]{$l_1$}}
\put(46.00,63.00){\makebox(0,0)[cc]{$l_2$}}
\put(32.00,65.00){\makebox(0,0)[cc]{$l_n$}}
\put(46.33,49.33){\makebox(0,0)[cc]{$C_1$}}
\put(25.67,55.00){\makebox(0,0)[cc]{$C_n$}}
\put(121.00,77.67){\makebox(0,0)[cc]{$A_0$}}
\put(96.33,96.00){\makebox(0,0)[cc]{$A_1$}}
\put(44.00,100.67){\makebox(0,0)[cc]{$A_n$}}
\put(9.33,32.33){\makebox(0,0)[cc]{$\scriptstyle{M}$}}
\put(8.33,73.67){\makebox(0,0)[cb]{$\scriptstyle{\mu_c(0)}$}}
\put(10.33,128.67){\makebox(0,0)[rb]{$\mu$}}
\put(44.33,78.67){\line(5,-6){3.00}}
\put(63.67,77.33){\line(-2,-1){6.67}}
\put(49.00,73.33){\circle{0.33}}
\put(51.33,72.33){\circle{0.33}}
\put(54.00,72.67){\circle{0.33}}
\put(29.00,79.00){\line(-5,-4){4.67}}
\end{picture}
\end{center}
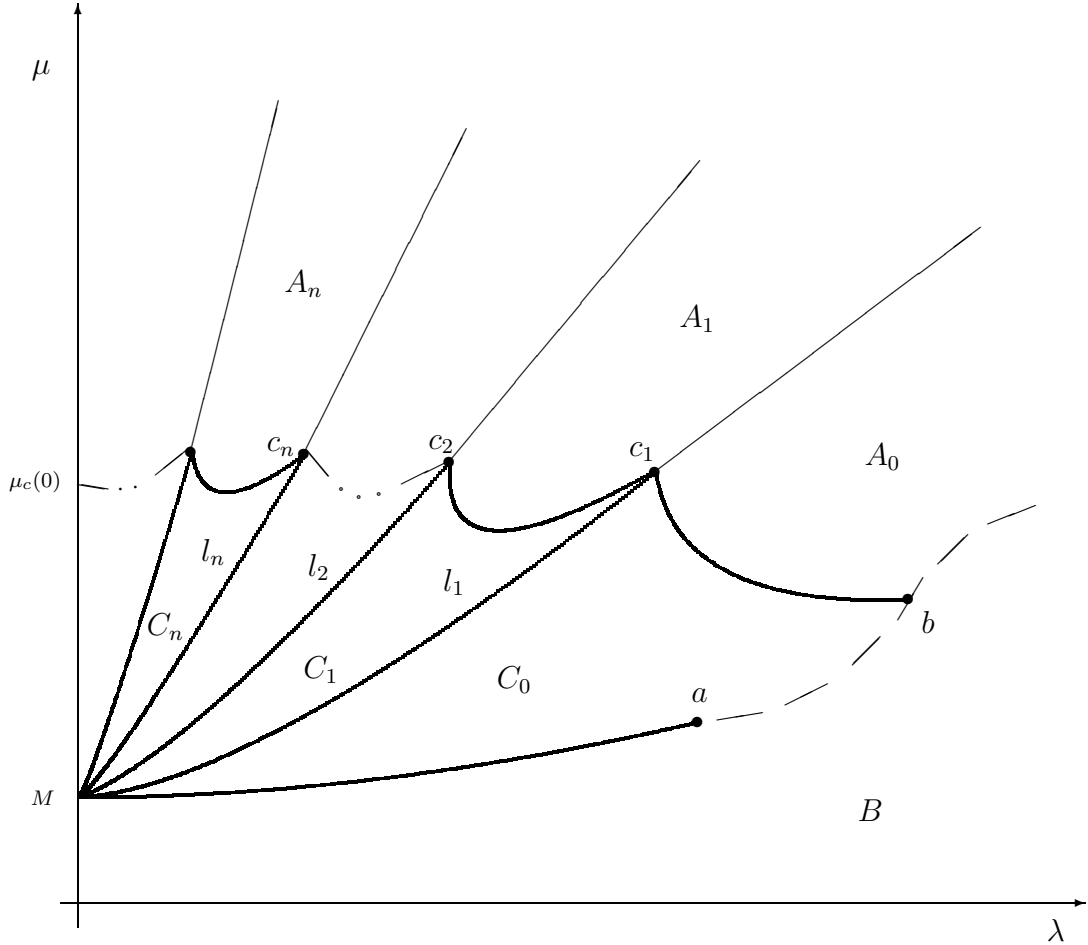
{\scriptsize FIG.3 Phase portrait of the $R^3\ts S^1$ NJL model
at $\mu\neq 0$ and $G_c<G<G_2$ ($\la=1/L$). Dashed lines are
critical curves of first-order phase transition, solid lines
correspond to second-order critical curves. Points $a$ and $b$
are tricritical points. There are cascades of massless symmetric
phases $A_k$ as well as massive phases $C_k$ ($k=0,1,2...$). The
line $Ma$ is $\mu=\Si_0(\la)$ and
$\mu_c(0)=2\pi\bar\la_0/\sqrt{6},$ where
$\f{\pi^2}{2G}-\f{\La^2}8\equiv -\f{\pi^2}6\bar\la^2_0$.}
\vspace{0.5cm}

\ul{Oscillations of the critical curve $\mu_{c}(\la)$.}
Recall that $\mu_{c}(\la)$ is the solution of the equation (13).
Evidently, inside an arbitrary region $\om_k$ (see (12)) this function has the form
\be
\mu_{c}(\la)\big|_{\om_k}\equiv\mu_{(k)}(\la)
=\frac{2\pi\{[ 6k(k+1) +1]\la^2+\bar\la^2_0\}}{6(2k+1)\la},
\ee
where $\bar\la_0$ is given in the caption to Fig.3. Hence,
\be
\mu_{c}(\la) =\mu_{(k)}(\la)~~~~~~~\mbox{at}~~~~~  t_{k+1}\leq\la\leq
t_k~~~ ,~~~~k=1,2,3,... ,
\ee
where $t_k$ is such a value of parameter $\la$, that the curve
$\mu_{c}(\la)$ crosses the line $\mu =2\pi k\la$, i.e. the right
boundary of $\om_k$:
\be
t_k=\frac{\bar\la_0}{\sqrt{6k^2-1}}.
\ee
Note, $\mu_{(k)}(t_k) =\mu_{(k-1)}(t_k)$, so the function $\mu_{c}(\la)$ (15)
is a continuous one at $\la> 0$. It follows also from (15) that
\[
\frac{d\mu_{(k-1)}(\la)}{d\la}\bigg|_{\la\to t_{k+}} =\f{\pi(2-6k)}{3(2k-1)} <0,
\]
\[
\frac{d\mu_{k}(\la)}{d\la}\bigg|_{\la\to t_{k-}} =\f{\pi(2+6k)}{3(2k+1)} >0.
\]
Last inequalities mean that at infinite set of points $t_k$
($k=1,2,....$) the function $\mu_{c}(\la)$ is not differentiated.
According to the above given definition, the critical curve $\mu_{c}(\la)$ 
oscillates at $\la\to 0$ or, equivalently, at $L\to\infty$ (see Fig.3).

Finally, let us present this oscillations of the $\mu_{c}(\la)$ in a manifest
form. We need the following Poisson's summation formula \ct{lan}:
\be
\sum^{\infty}_{n=0}\alpha_n\Phi (n)
=2\sum^{\infty}_{k=0}\alpha_k
\int\limits^{\infty}_{0}\Phi (x)
\cos (2\pi kx)dx,
\ee
where $\al_n=2-\delta_{n0}$. Using it in equation (13), one can easily find
at $\la\to 0$:
\be
\mu_c(\la)\approx\f{2\pi\bar\la_0}{\sqrt{6}}\left\{1+\f{3\la^2}{\pi^2\bar\la^2_0}
\sum_{n=1}^\infty \f {\cos(n\pi\bar\la_0L/\sqrt{6})}{n^2}\right\},
\ee
 From (18) it
follows that $\mu_{c}(\la)$ has an oscillating part, which
oscillates at $L\to\infty$ with frequency $\bar\la_0/(2\sqrt{6})$.

\ul{Oscillations of the fermionic condensate.}
Fermionic condensate is defined as $<\bar\psi\psi>$, and in NJL
model it equals to $<\Si>$. Since the last quantity is the
global minimum point $\Si(\mu,\la)$ of an effective potential,
we should study nontrivial solution of the stationary equation
(11). Detailed analysis of $\Si(\mu,\la)$ was done in \ct{2},
and this quantity at $M<\mu<\mu_c(0)$ and at $\la\to 0$
($L\to\infty$) behaves as
\be
\Si(\mu,\la)=m(\mu)+\f{\la^2\sqrt{\mu^2-m^2(\mu)}}{\mu f'(m(\mu))}
\sum_{n=1}^\infty \f{\cos(n\sqrt{\mu^2-m^2(\mu)}~L)}{n^2}+o(\la^2),
\ee
where $m(\mu)$ equals to $\Si(\mu,0)$, $f'(m)$ is the derivative
of the function
\be
f(m)\equiv F(m)+\f\mu 4\sqrt{\mu^2-m^2}-
\f{m^2}4\ln \left (\f{\mu+\sqrt{\mu^2-m^2}}m\right ),
\ee
and $F(\Si)$ is given in (6). It is clear from (19) that $\Si(\mu,\la)$
has oscillating part, which oscillates at $L\to\infty$ with frequency
$\sqrt{\mu^2-m^2(\mu)}/(2\pi)$.

\ul{Oscillations of the particle density.} Suppose, $M<\mu<\mu_c(0)$.
Then the thermodynamic potential (TDP) $\Om(\mu,\la)$ of the NJL
system is equal to the value of its effective potential at the
global minimum point $\Si(\mu,\la)$, i.e. $\Om(\mu,\la)$$=V_{\mu
L}(\Si(\mu,\la)$. It is well-known that thermodynamic potential
defines the particle density $n(\mu,\la)$ through the relation:
$n(\mu,\la)$$=-\p\Om(\mu,\la)/\p\mu$. Hence,
\[
n(\mu,\la)=-\Biggl\{
\frac{\partial V_{\mu L}(\Sigma)}{\partial\mu}+
\frac{\partial V_{\mu L}(\Sigma)}{\partial\Sigma}
\frac{\partial \Sigma}{\partial\mu}
\Biggr\}
\Biggl|_{\Sigma=\Sigma(\mu,\la)}\Biggr.
\]
\be
=\f{\la}{2\pi}\sum^{\infty}_{n=0}\al_n
\Theta (\mu - \sqrt{\Sigma^2(\mu,\la) + (2\pi\lambda n)^2}~)
(\mu^2 -\Sigma^2(\mu,\la) - (2\pi\lambda n)^2~).
\ee
Using in (21) the Poisson's summation formula (17) \ct{2}, we see that
at $\mu=const$ and $L\to\infty$ 
\[
n(\mu,\la)=\f{(\mu^2-m^2(\mu))^{3/2}}{3\pi^2}
+\la^2\left [\f{m(\mu)(\mu^2-m^2(\mu))}{\mu f'(m(\mu))}-2\sqrt{\mu^2-m^2(\mu)}\right ]\cdot
\]
\be
\cdot\sum^{\infty}_{n=1}\f{\cos(n\sqrt{\mu^2-m^2(\mu)}L)}{\pi^2n^2}+o(\la^2),
\ee
where $m(\mu)$ is the fermion mass at $\la=1/L=0$ (see (19)),
$f(m)$ is defined in (20). From (22) one can easily see that particle
density in the ground state of the NJL model oscillates with frequency
$\sqrt{\mu^2-m^2(\mu)}/(2\pi)$.

\ul{Oscillations of the pressure.} Let us suppose that $\mu>\mu_c(0)$,
i.e. for sufficiently large values of $L$ the global minimum point
of the effective potential equals to zero. In this case the TDP of the model
equals to $V_{\mu L}(0)$. Then, at $L\to\infty$ the TDP $\Om (\mu,\la)$
oscillates with frequency $\mu/(2\pi)$, because it looks like \ct{2}:
\[
\Om(\mu,\la)=V_L(0)-\f{\mu^4}{12\pi^2}-\sum^{\infty}_{k=0}\left
[\f{4\la^4}{\pi^2k^4}-\f{2\mu\la^3}{\pi^2k^3}\sin(\mu
kL)-\f{4\la^4}{\pi^2k^4}\cos(\mu kL)\right ].  
\]
In our case the pressure in the system is defined
as $p=-\p(L\Om)/\p L$. Using the above expression for $\Om(\mu,\la)$,
 we see that the pressure in the vacuum of the NJL model also oscillates with
frequency $\mu/(2\pi)$.

One can interpret the case under consideration as the ground
state of the NJL system, located between two parallel plates with
periodic boundary conditions. The force which acts on each of
plates is known as  generalized  Casimir force. Evidently, this
force is propotional to the pressure in the ground state of the
system. Hence, at nonzero chemical potential the Casimir force
of the constrained fermionic vacuum is oscillates at
$L\to\infty$.

\section*{3. Phase structure of the NJL model at $\mu\neq 0$ and in the
presence of external magnetic field}

In the present section we shall study the magnetic properties of the
NJL vacuum. At $\mu =0$ this problem was considered in
\ct{klev,gus}.  It was shown in \ct{klev} that at $G>G_c$ the
chiral symmetry is spontaneously broken for arbitrary values of
external magnetic field $H$, and even for $H=0$. At $G<G_c$ the
NJL system has a symmetric vacuum at $H=0$. However, if the external
(arbitrary small) magnetic field is switched on, then for all
$G\in (0,G_c)$ one has a spontaneous breaking of initial
symmetry \ct{gus}. This is a so called effect of dynamical
chiral symmetry breaking catalysis by external magnetic field.

The brief history of this effect is the following.  First of all
such property of external magnetic field was discovered in (2+1)
- dimensional Gross-Neveu (3DGN) model \ct{3,4}. At $H=0$ there
exist two phases in 3DGN model: one of which is a massless
chirally invariant phase ($G<G_c$), and another one is a massive phase with
spontaneously broken chiral symmetry ($G>G_c$). However, for each value of
$H\neq 0$ as well as for all values of the bare coupling
constant $G>0$ the symmetric phase is absent in a 3DGN theory,
and chiral symmetry is broken down \ct{3,4} \footnote{The
consideration in \ct{4} is performed in terms of parameter $g$,
such that $\f 1g=\f 1{g(m)}-\f {2m}\pi$, where $m$ is a normalization point, $g(m)$ is renormalized
coupling constant. The connection between $g$ and $G$ was
established in \ct{5}: $\f 1g=\f 1G-\f 1G_c$. So, at $g<0$ and
$g>0$ one has $G>G_c$ and $G<G_c$, respectively.}.  Of course,
in \ct{3,4} a special consideration was done for the case
$G<G_c$, where the magnetic field induces the DCSB even for
weakest attractive interaction between fermions (the magnetic
catalysis of DCSB).  It turns out, that external chromomagnetic
field is also a magnetic catalyst of DCSB \ct{5}. The influence
of temperature and chemical potential on this effect was studied
in \ct{4}-\ct{7}, where the restoration of chiral symmetry at
sufficienly large values of $T$ and $\mu$ was predicted.  Later,
in \ct{gus1} the explanation of this phenomenon on the basis of
dimensional reduction mechanism was found in the framework of
3DGN model.  The magnetic catalysis takes place in
four-dimensional NJL model \ct{gus,ebert,bab} as well as in
other theories, and now it is under intensive consideration
(see, e.g. \ct{muta,par,mir} and references therein).

(Authors of \ct{bab,mir} declare that in our paper \ct{4} only
the "fact that external magnetic field enhances a fermion
dynamical mass" was established. Hence, they assert that in
\ct{4} only the case $G>G_c$ was considered. In reality, in
\ct{3,4,5} the action of external (chromo-)magnetic field on the
3DGN model was studied for arbitrary values of bare coupling
constant. The spontaneous breakdown of chiral symmetry was
found there for all $G\in(0,\infty)$, including the case
$G<G_c$, and even the case of arbitrary small values of $G>0$
(the magnetic catalysis of DCSB).)

In the present section we continue the investigatuion of
magnetic catalysis effect and this time turn to the
consideration of the four-dimensional NJL model at $H,\mu\neq 0$.

\subsubsection*{3.1 Magnetic catalysis at $\mu\neq 0$}

Let us recall some aspects of the problem at $\mu =0$.  Using a
well-known proper-time method \ct{sch} or a momentum-space
calculations \ct{dit}, one can find the effective potential
$V_H(\Si)$ of the NJL model at $H\neq 0$:
\[
V_H(\Si)=\f{\Si^2}{2G}+\f{eH}{8\pi^2}
\int_{0}^{\infty} \f{ds}{s^2} \exp (-s\Si^2)~\coth(eHs).
\]
After identical transformations we have
\be
V_H(\Si)=V_0(\Si)+\ti V_H(\Si)+Z(\Si),
\ee
where
\ba
V_0(\Si)&=&\f{\Si^2}{2G}+\f{1}{8\pi^2}\int_{0}^{\infty}
\f{ds}{s^3} \exp (-s\Si^2),\nn \\
Z(\Si)&=&\f{(eH)^2}{24\pi^2}
\int_{0}^{\infty} \f{ds}{s^3} \exp (-s\Si^2),\nn \\
\ti V_H(\Si)&=&\f{1}{8\pi^2}
\int_{0}^{\infty} \f{ds}{s^3} \exp (-s\Si^2)\left [~(eHs)\coth(eHs)
-1-\f{(eHs)^2}3~\right ].
\ea
The potential $V_0(\Si)$ in (24) up to infinite additive constant is
equal to the function (4). Hence, the UV-regularized expression for it
looks like (5).

The function $Z(\Si)$ is also an UV-divergent one, so we need to regularize it:
\ba
Z(\Si)&=&\f{(eH)^2}{24\pi^2}\int_{0}^{\infty} \f{ds}{s}( \exp (-s\Si^2)
-\exp (-s\La^2))+\f{(eH)^2}{24\pi^2}
\int_{0}^{\infty} \f{ds}{s} \exp (-s\La^2) \nn \\
&=&-\f{(eH)^2}{24\pi^2}\ln\f{\Si^2}{\La^2}+\f{(eH)^2}{24\pi^2}
\int_{0}^{\infty} \f{ds}{s} \exp (-s\La^2).
\ea
The infinite last term in (25) contributes to the
renormalization of an electric charge and magnetic field as
well, similar as it occures in quantum electrodynamics \ct{sch}.
 
The potential $\ti V_H(\Si)$ in (25) has no UV divergences, so it is
easily calculated with the help of a table of integrals
\ct{prud}. The final expression for $V_H(\Si)$ is:
\be
V_H(\Si)=V_0(\Si)-\f{(eH)^2}{2\pi^2}\Bigl\{\zeta '(-1,x)-\f 12[x^2-x]\ln x
+\f{x^2}4\Bigr\},
\ee
where $x=\Si^2/(2eH)$, $\zeta (\nu,x)$ is the generalized Riemann
zeta-function and $\zeta'(-1,x)$$=d\zeta(\nu,x)/d\nu|_{\nu=-1}$.
 The global minimum point of this function is
the solution of the stationary equation: 
\be
\f {\p}{\p\Si}V_H(\Si)=\f{\Si}{4\pi^2}\{F(\Si)-I(\Si)\}=0,
\ee
where $F(\Si)$ is given in (6), and 
\ba
I(\Si)&=&2eH\{\ln\Ga(x)-\f 12\ln(2\pi)+x-\f 12(2x-1)\ln x\}\nn \\
&=& \int_{0}^{\infty} \f{ds}{s^2} \exp (-s\Si^2)[~eHs\coth(eHs)-1]
\ea
For arbitrary fixed values of $H,G$ there is only one nontrivial
solution $\Si_0(H)$ of equation (27), which is the global
minimum point of $V_H(\Si)$. 

Hence, at $G<G_c$ and $H=0$ the NJL vacuum is chirally symmetric one, but
arbitrary small value of external magnetic field $H$ induces the DCSB, and
fermions acquire nonzero mass $\Si_0(H)$ (the effect of magnetic
catalysis of DCSB).

In the present paper we shall consider only the case
$G<G_c$. Therein, $\Si_0(H)$ is a monotonically increasing function versus $H$.
Apart from, at $H\to\infty$  
\be
\Si_0(H)\approx\f {eH}\pi \sqrt{\f G{12}}
\ee
and at $H\to 0$
\be
\Si_0^2(H)\approx\f {eH}\pi\exp\{-\f 1{eH}(\f {4\pi^2}G-\La^2)\}
\ee

Now let us consider a more general case, when $H\neq 0,\mu\not = 0$.
In one of our prevoius papers \ct{7} the effective potential of
a 3DGN model at nonzero $H$, $\mu$ and $T$ was obtained. Similarly, one
can find an effective potential in the NJL model at $H,T,\mu\neq 0$:
\be
V_{H\mu T}(\Si)=V_H(\Si)-\f{TeH}{4\pi^2}\sum_{k=0}^{\infty} \al_k
\int_{-\infty}^\infty dp\ln\left\{\left [1+\exp^{-\beta(\ve_k+\mu)}\right ]
\left [1+\exp^{-\beta(\ve_k-\mu)}\right ]\right\},
\ee
where  $\beta=1/T$, $\al_k=2-\del_{0k}$, $\ve_k=\sqrt{\Si^2+p^2+2eHk}$, and
a function $V_H(\Si)$ is given in (26). Tending a temperature in (31)
to the zero, we have the effective potential of NJL model at $H,\mu\neq 0$:
\be
V_{H\mu}(\Si)=V_H(\Si)-\f{eH}{4\pi^2}\sum_{k=0}^{\infty} \al_k
\int_{-\infty}^\infty dp (\mu-\ve_k)\theta (\mu-\ve_k),
\ee
which can be easily transformed to the form
\be
V_{H\mu}(\Si)=V_H(\Si)-\f{eH}{4\pi^2}\sum_{k=0}^{\infty} \al_k
\theta (\mu-s_k)\Biggl\{\mu\sqrt{\mu^2-s_k^2}-s_k^2\ln\left
[\f {\mu+\sqrt{\mu^2-s_k^2}}{s_k}\right ]\Biggr\},
\ee
where $s_k=\sqrt{\Si^2+2eHk}$. Finally, let us present the stationary 
equation for the potential (33):
\be
\f {\p}{\p\Si}V_{H\mu}(\Si)=\f{\Si}{4\pi^2}\Biggl\{F(\Si)-I(\Si)
+2eH\sum_{k=0}^{\infty} \al_k \theta (\mu-s_k)\ln\left
[\f {\mu+\sqrt{\mu^2-s_k^2}}{s_k}\right ]\Biggr\}=0,
\ee
In order to get a phase portrait of the model one should find a one-to-one
correspondence between points of $(\mu,H)$ plane and global minimum points
of the function (33), i.e. we need to solve equation (34), find a global
minimum $\Si (\mu,H)$ for potential (33), to study properties of $\Si (\mu,H)$
versus $(\mu,H)$.

In order to greatly simplify this problem let us divide the plane $(\mu,H)$
into a set of regions $\om_k$:
\be
(\mu, H)=\bigcup\limits^{\infty}_{k=0}\omega_k;\,\,
\omega_k=\{(\mu, H):2eHk\leq \mu^2\leq 2eH (k+1)\}.
\ee
In the region $\om_0$ only a first term is nonzero from a series in (34-35).
So, one can find that for the points $(\mu,H)\in\om_0$ which are above
the line $L$=$\{(\mu,H):\mu=\Si_0(H)\}$, the global minimum is at the point
$\Si=0$. Just under the curve $L$ the point $\Si=\Si_0(H)$ is a local minimum
of the potential (33), and $\Si=\Si_0(H)$ transforms to the global
minimum when $(\mu,H)$ lies under the critical curve of the first order
phase transitions $\mu=\mu_c(H)$, which is defined by the following equation:
\be
V_{H\mu}(0)=V_{H\mu}(\Si_0(H)).
\ee
In the region $\om_0$ one can easily solve this equation:
\be
\mu_c(H)=\f {2\pi}{\sqrt{eH}} [V_H(0)-V_H(\Si_0(H))]^{1/2}.
\ee

Hence, we have shown that at $\mu>\mu_c(H)$ ($G<G_c$) there is a massless
symmetric phase of the NJL model (numerical investigations of (33-34)
gives us the zero global minimum point for the potential $V_{H\mu}(\Si)$
in other regions $\om_1,\om_2,...$ as well). The external magnetic
field ceases to induce the DCSB at $\mu>\mu_c(H)$
(or at sufficiently small values of magnetic field $H<H_c(\mu)$, where
$H_c(\mu)$ is the inverse function to $\mu_c(H)$).
But, under the critical curve (37) (or at $H>H_c(\mu)$) due to a
presence of external magnetic field the chiral symmetry is
spontaneously broken. Here magnetic field induces dynamical
fermion mass $\Si_0(H)$, which is not $\mu$-dependent value.

At last, we should remark that in the NJL model 
the magnetic catalysis effect takes place only in the
phase with zero particle density, i.e. at $\mu<\mu_c(H)$. If
$\mu>\mu_c(H)$ we have symmetric phase with nonzero particle
density, but here the magnetic field can not induces DCSB.

\subsubsection*{3.2 Magnetic oscillations}

In a previous case we have shown that points $(\mu,H)$,
which are above critical curve $\mu=\mu_c(H)$, correspond to the 
chirally symmetric ground state of the NJL model. One fermionic excitations
of this vacuum have zero masses. At first sight, properties of this 
simmetric vacuum are slightly varied, when parameters $\mu$ and $H$ are
changed. However, it is not so and in the region $\mu>\mu_c(H)$ 
we have infinitly many massless symmetric phases of the theory
as well as a variety of critical curves of the second order
phase transitions. On the experiment this cascade of phases is
identified with oscillations of such physical quantities as
magnetization and particle density.  Let us prove it.

It is well-known that a state of the thermodynamic equilibrium ($\equiv$
the ground state) of arbitrary quantum system is described by the 
thermodynamic potential (TDP) $\Om$, which is
a value of the effective potential in its global minimum point.
In the case under consideration the TDP $\Om(\mu,H)$ 
at $\mu>\mu_c(H)$ has the form
\[
\Om(\mu,H)\equiv V_{H\mu}(0)=V_H(0)-
\]
\be
-\frac{eH}{4\pi^2}\sum^{\infty}_{k=0}\alpha_k\theta
(\mu-\e_k)\{~\mu\sqrt{\mu^2-\e_k^2}-\e_k^2\ln[(\sqrt{\mu^2-\e_k^2}
+\mu)/\e_k]~\},
\ee
where $\e_k=\sqrt{2eHk}$. We shall use the following criterion
of the phase transitions: if at least one first (second) partial
derivative of $\Om(\mu,H)$ is a discontinuous function at some
point, then it is a point of the first (second) order
phase transition.

Using this criterion let us show that boundaries of $\om_k$ regions (35),
i.e. lines $l_k=\{(\mu,H):\mu=\sqrt{2eHk}\}$ ($k=1,2,...$), 
are critical lines of second order phase transitions.
In arbitrary region $\om_k$ the TDP (38) has the form:
\[
\Om(\mu,H)\big |_{\om_k}\equiv \Om_k=V_H(0)-
\]
\be
-\frac{eH}{4\pi^2}\sum^{k}_{i=0}\alpha_i \theta (\mu-\epsilon_i)
\Biggl\{~\mu\sqrt{\mu^2-\e_k^2}-\e_i^2\ln\left
[\f {(\sqrt{\mu^2-\e_i^2}+\mu)}{\e_i}\right ]\Biggr\}.
\ee
From (39) one can easily find
\be
\f {\p\Om_k}{\p\mu}\bigg |_{(\mu,H)\to l_{k+}}-
\f {\p\Om_{k-1}}{\p\mu}\bigg |_{(\mu,H)\to l_{k-}}=0,
\ee
as well as:
\be
\f {\p^2\Om_k}{(\p\mu)^2}\bigg |_{(\mu,H)\to l_{k+}}-
\f {\p^2\Om_{k-1}}{(\p\mu)^2}\bigg |_{(\mu,H)\to l_{k-}}=
-\f {eH\mu}{2\pi^2\sqrt{\mu^2-\e_k^2}}\bigg |_{\mu\to \e_{k+}}\ra -\infty.
\ee
The equality (40) means that the first derivative $\p\Om/\p\mu$
is a continuous function at all lines $l_k$. However, the second
derivative $\p^2\Om/(\p\mu)^2$ has an infinite jump at each line
$l_k$ (see (41)), so these lines are critical curves of second
order phase transitions.  (Similarly, one can prove the
discontinuity of $\p^2\Om/(\p H)^2$ and $\p^2\Om/\p\mu\p H$ at
all lines $l_n$.)

Let the chemical potential be fixed, i.e. $\mu=const$. Then on
the plane $(\mu,H)$ we have a line, which crosses critical lines
$l_1,l_2,...$ at points $H_1,H_2,...$ correspondingly.  The
particle density $n$ and the magnetization $m$ of any
thermodynamic system are defined by the TDP in a following way:
$n=-\p\Om/\p\mu$, $m=-\p\Om/\p H$.  At $\mu=const$ these
quantities are continuous functions over external magnetic field
only, i.e. $n\equiv n(H),~m\equiv m(H)$. We know that all second
derivatives of $\Om(\mu,H)$ are discontinuous at every critical
line $l_n$. So, functions $n(H)$ and $m(H)$, continuous at the
interval $H\in (0,\infty)$, have derivatives broken at infinite set of 
points $H_1,...,H_k,...$. According to the definition given in
section 2.1 particle density and magnetization oscillate at
$H\to 0$.

In order to present oscillating parts of $n(H)$ and $m(H)$ in a manifest 
analitic form, we shall use the technique elaborated in \ct{as}, where
a manifest analitic expression was found for oscillating part of
$\Om(\mu,H)$ for free relativistic electron - positron gas. This
technique may be used without any difficulties in our case as well.

Hence, one can rewrite the TDP (38) in the following form:
\be
\Om(\mu,H)=\Om_{mon}(\mu,H)+\Om_{osc}(\mu,H),
\ee
where ($\nu=\mu^2/(eH)$):
\be
\Omega_{mon}=V_H(0)-\f {\mu^4}{12\pi^2}-\frac{(eH)^2}{4\pi^3}
\int\limits^{\nu}_{0}dy \sum^{\infty}_{k=1}\frac{1}{k}P(\pi ky),
\ee
\be
\Omega_{osc}=\frac{\mu}{4\pi^{3/2}}
\sum^{\infty}_{k=1}
\left (\frac{eH}{\pi k}\right)^{3/2}
[Q(\pi k \nu)\cos (\pi k\nu+\pi/4)+
P(\pi k \nu)\cos (\pi k\nu -\pi/4)].
\ee
(To find (43-44) it is sufficient to tend the electronic mass to zero
in the formula (19) from \ct{as}.) Functions $P(x)$ and $Q(x)$ in (43-44)
are connected with Fresnel's integrals $C(x)$ ¨ $S(x)$ \ct{15}:
\by
C(x)&=&\frac{1}{2}+\sqrt{\frac{x}{2\pi}}
[P(x)\sin x +Q(x)\cos x] \\
S(x)&=&\frac{1}{2}-\sqrt{\frac{x}{2\pi}}
[P(x)\cos x -Q (x)\sin x].
\ey
They have at $x\to\infty$ following asimptotics \ct{15}:
\[
P(x)=x^{-1}-\frac{3}{4}x^{-3}+...,~~~Q(x)=-\frac{1}{2}x^{-2}+\frac{15}{8}x^{-4}+...
\]
The formula (44) presents the exact oscillating part of the TDP (38) for 
the NJL model at $G<G_c$. Since in the present case the TDP is propotional
to the pressure of the system, one can conclude that pressure in the NJL
model oscillates, when $H\to 0$.  It follows from (44) that
frequency of oscillations at large values of a parameter
$(eH)^{-1}$ equals to $\mu^2/2$. Then, starting from (44) one can
easily find manifest expression for oscillating parts of $n(H)$
and $m(H)$. These quantities oscillates at $H\to 0$ with the
same frequency $\mu^2/2$ and have a rather involved form, so we
do not present it here.
\vspace{0.5cm}

The author is grateful to Prof. U. Heinz and to the Organizing Committee
of the "5th International Workshop on TFT and Their Application"
for kind invitation to attend this Workshop as well as to the Deutsche Forschungsgemeinschaft
for financial support.
This work was supported in part by the Russian Fund for Fundamental 
Research, project 98-02-16690.

\vspace*{0.5cm}




\begin{thebibliography}{99}

\bibitem{1}
A.S. Vshivtsev and K.G. Klimenko, JETP Lett. {\bf 64}, 338 (1996); hep-ph/9701288;\\
A.S. Vshivtsev, V.Ch. Zhukovskii and K.G. Klimenko, JETP {\bf 84}, 1047 (1997).
\bibitem{2}
A.S. Vshivtsev,  A.K. Klimenko and K.G. Klimenko, Phys. Atom. Nucl. {\bf 61}, 479 (1998);\\
M.A. Vdovichenko, A.S. Vshivtsev and K.G. Klimenko, preprint IFVE 97-59, 
Protvino (1997) (in russian); M.A. Vdovichenko, A.S. Vshivtsev and K.G. Klimenko,
JETP (1998) (to be published).
\bibitem{nam}
Y. Nambu and G. Jona-Lasinio, Phys. Rev. {\bf 122}, 345 (1961).
\bibitem{vaks}
V.G. Vaks, A.I. Larkin, Sov. Phys. JETP {\bf 13}, 192; 979 (1961);\\
B.A. Arbuzov, A.N. Tavkhelidze, R.N. Faustov, Sov. Phys. Dokl. {\bf 6}, 598 (1962). 
\bibitem{kaw}
S. Kawati and H. Miyata, Phys. Rev. {\bf D 23}, 3010 (1981);\\
J. Fuchs, Z.Phys. {\bf C 22}, 83 (1984);\\
V. Bernard, U.-G. Meissner and I. Zahed, Phys. Rev. {\bf D 36}, 819 (1987);\\
Chr.V. Christov and K. Goeke, Acta Phys. Pol. {\bf B 22}, 187 (1991);\\
D. Ebert, Yu.L. Kalinovsky, L. M\"unchow and M.K. Volkov, Int. J. Mod. Phys.
{\bf A 8}, 1295 (1993).
\bibitem{klev}
S.P. Klevansky and R.H. Lemmer, Phys. Rev. {\bf D 39}, 3478 (1989).
\bibitem{eb}
D. Ebert and M.K. Volkov, Phys. Lett. {\bf B 272}, 86 (1991);\\
I.A. Shovkovy and V.M. Turkowski, Phys. Lett. {\bf B 367}, 213 (1995).
\bibitem{gus}
V.P. Gusynin, V.A. Miransky and I.A. Shovkovy, Phys. Lett. {\bf B 349}, 477 (1995).
\bibitem{In}
T. Inagaki, T. Muta and S.D. Odintsov, Mod. Phys. Lett. 
{\bf A 8}, 2117 (1993);\\
E. Elizalde, S. Leseduarte and S.D. Odintsov, Phys. Rev. 
{\bf D 49}, 5551 (1994);\\E. Elizalde, S. Leseduarte and S.D. Odintsov,
Phys. Lett. {\bf B 347}, 33 (1995);\\
H. Forkel, Phys. Lett. {\bf B 280}, 5 (1992); Nucl. Phys. {\bf A 581}, 557 (1995);\\
D.K. Kim and K.G. Klimenko, J. Phys. {\bf A 31}, 5565 (1998).
\bibitem{muta}
T. Inagaki, T. Muta and S.D. Odintsov, Progr. Theor. Phys. Suppl.
{\bf 127}, 93 (1997).
\bibitem{od}
D.M. Gitman, S.D. Odintsov and Yu.I. Shil'nov, Phys. Rev. {\bf D 54}, 2964 (1996);\\
B. Geyer, L.N. Granda and S.D. Odintsov, Mod. Phys. Lett. {\bf
A 11}, 2053 (1996);\\ T. Inagaki, S.D. Odintsov and Yu.I. Shil'nov,
KOBE-TH-97-02, hep-th/9709077.
\bibitem{dol}
L. Dolan and R. Jackiw, Phys. Rev. {\bf D 9}, 3320 (1974).
\bibitem{kim}
S.K. Kim, W. Namgung, K.S. Soh and J.H. Yee, Phys. Rev. {\bf D 36}, 3172 (1987);\\
D.Y. Song, Phys. Rev. {\bf D 48}, 3925 (1993);\\
D.K. Kim, Y.D. Han and I.G. Koh, Phys. Rev. {\bf D 49}, 6943 (1994);\\ 
D.K. Kim and I.G. Koh, Phys. Rev. {\bf D 51}, 4573 (1995);\\
K.G. Klimenko, A.S. Vshivtsev and B.V. Magnitskii, JETP Lett. {\bf 61}, 871 (1995).
\bibitem{per}
Yu.B. Rumer and M.Sh. Ryvkin, {\it Thermodynamics, Statistical Physics and Kinetics},
Nauka, Moscow (1977);\\D. Persson and V. Zeitlin, Phys. Rev. {\bf D 51}, 2026 (1995);\\
J.O. Andersen and T. Haugset, Phys. Rev. {\bf D 51}, 3073 (1995);\\
V.Ch. Zhukovskii, T.L. Shoniya and P.A. Eminov, JETP {\bf 80}, 158 (1995).
\bibitem{as}
A.S. Vshivtsev and K.G. Klimenko, JETP {\bf 82}, 514 (1996); 
\bibitem{lan}
L.D. Landau and E.M. Lifshitz, {\it Statistical Physics},
Nauka, Moscow (1976).
\bibitem{3} 
K.G. Klimenko, Theor. Math. Phys. {\bf 89}, 1161 (1992).
\bibitem{4}
K.G. Klimenko, Z.Phys. {\bf C 54}, 323 (1992).
\bibitem{5}
K.G. Klimenko, A.S. Vshivtsev and B.V. Magnitsky, Nuovo Cim.
{\bf A 107}, 439 (1994);\\K.G. Klimenko, A.S. Vshivtsev and B.V. Magnitsky,
Theor. Math. Phys. {\bf 101}, 1436 (1994);\\K.G. Klimenko, A.S. Vshivtsev and B.V. Magnitsky,
Phys. Atom. Nucl. {\bf 57}, 2171 (1994);\\
K.G. Klimenko, A.S. Vshivtsev and B.V. Magnitsky, in Proc. of
"3rd Workshop on Thermal Field Theories and Their Applications",
World Scientific, Singapore (1994).
\bibitem{6}
K.G. Klimenko, Theor. Math. Phys. {\bf 90}, 1 (1992).
\bibitem{7}
A.S. Vshivtsev, K.G. Klimenko and B.V. Magnitsky,
Theor. Math. Phys. {\bf 106}, 319 (1996).
\bibitem{gus1}
V.P. Gusynin, V.A. Miransky and I.A. Shovkovy, Phys. Rev.
Lett. {\bf 73}, 3499 (1994).
\bibitem{ebert}
D. Ebert and V.Ch. Zhukovsky, Mod. Phys. Lett. {\bf A 12}, 2567 (1997).
\bibitem{bab}
A.Yu. Babansky, E.V. Gorbar and G.V. Shchepanyuk, Phys. Lett. {\bf B 419}, 272 (1998).
\bibitem{par}
R.R. Parwani, Phys. Lett. {\bf B 358,} 101 (1995);\\
W. Dittrich and H. Gies, Phys. Lett. {\bf B 392}, 182 (1997);\\
I.A. Shushpanov and A.V. Smilga, Phys. Lett. {\bf B 402}, 351 (1997);\\
V.P. Gusynin and I.A. Shovkovy, Phys. Rev. {\bf D 56}, 5251 (1997);\\
V.P. Gusynin, hep-ph/9709339; I.A. Shovkovy, hep-ph/9709340;\\
D.-S. Lee, C.N. Leung and Y.J. Ng, Phys. Rev. {\bf D 57}, 5224 (1998);\\
E. Elizalde, Yu.I. Shil'nov, and V.V. Chitov, Class. Quant. Grav. {\bf 15}, 735 (1998);\\
E. Elizalde, S.P. Gavrilov, S.D. Odintsov and Yu.I. Shil'nov, hep-ph/9807368.
W.V. Liu, cond-mat/9808134.
\bibitem{mir}
V.A. Miransky, hep-th/9805159.
\bibitem{sch}
J. Schwinger, Phys. Rev. {\bf 82}, 664 (1951).
\bibitem{dit}
M.R. Brown and M.J. Duff, Phys. Rev. {\bf D 11}, 2124 (1975);\\
W. Dittrich, Fortsh. Phys. {\bf 26}, 289 (1978).
\bibitem{prud}
A.P. Prudnikov, Yu.A. Brychkov and O.I. Marichev, {\it Integrals and Series,}
Gordon and Breach, New York (1986).
\bibitem{15}
H. Bateman and A. Erdeyi, {\it Higher Transcendental Functions,}
McGrawHill, New York (1953).


\end{thebibliography}
\end{document}